\documentclass[useAMS,usenatbib]{mn2e}

\usepackage{graphicx}
\usepackage{bm}
\usepackage{psfrag}
\usepackage{mathrsfs}
\usepackage{amssymb,amsmath}
\newcommand{\nodata}{...}

\newcommand{\apj}{ApJ}

\newcommand{\apjl}{ApJL}
\newcommand{\aap}{A{\&}A}

\newcommand{\mnras}{MNRAS}

\newcommand{\araa}{ARAA}

\newcommand{\nat}{Nature}

\newcommand{\prd}{PRD}
\newcommand{\physrep}{Phys. Rep.}

\newcommand{\be}{\begin{eqnarray}}
\newcommand{\ee}{\end{eqnarray}}

\title[Radio-wave propagation and precision pulsar timing]{Modelling  and mitigating refractive propagation effects  in precision pulsar timing observations}

\author[R.~M.~Shannon \& J.~M.~Cordes ]{ R.~M.~Shannon\thanks{E-mail: ryan.shannon@csiro.au, cordes@astro.cornell.edu}$^{1,2}$,  J.~M.~Cordes$^{3}$\\
$^1$ CSIRO Astronomy and Space Science, Box 76 Epping, NSW, 1710, Australia \\
$^2$ International Centre for Radio Astronomy Research, Curtin University, Bentley, WA 6102, Australia \\
$^3$ Astronomy Department, Cornell University, Ithaca, NY, 14853, USA }

\begin{document}

\date{Accepted XX ; in original form \today}

\pagerange{1--16} \pubyear{2015}

\maketitle

\begin{abstract}
To obtain the most accurate pulse arrival times from radio pulsars, it is necessary to correct or mitigate the effects of the propagation of radio waves through the warm and ionised interstellar medium. 
We examine both the strength of propagation effects associated with large-scale electron-density variations and the methodology used to estimate infinite-frequency arrival times.
Using simulations of  two-dimensional phase-varying screens, we assess the strength and non-stationarity of timing perturbations associated with  large-scale density variations.
We identify additional  contributions to arrival times that are stochastic in both radio frequency and time and therefore not amenable to correction solely using times of arrival.
We attribute this to the frequency dependence of the trajectories of the propagating radio waves.  We find that this limits the efficacy of  low-frequency (metre-wavelength) observations.  
Incorporating low-frequency pulsar observations into precision timing campaigns is increasingly problematic for pulsars with larger dispersion measures. 

\end{abstract}



\section{Introduction}

The pulsar-timing technique has enabled many studies fundamental to physics and astrophysics, including  precise tests of  general relativity \cite[][]{1975ApJ...195L..51H,2006Sci...314...97K},  constraints on the equation of state of dense nuclear matter \cite[][]{2007PhR...442..109L,2010Natur.467.1081D}, and the discovery of the first planetary mass objects outside of the solar system \cite[][]{1992Natur.355..145W}. 

One possible application of the pulsar-timing technique is the  detection of gravitational waves (GWs)
 and the characterisation of GW-emitting sources.  
GWs  passing through the solar neighbourhood manifest as  variations in arrival times in an array of (MSPs)  \cite[a pulsar timing array, PTA; ][]{1979ApJ...234.1100D,1983ApJ...265L..39H,1990ApJ...361..300F} that have a quadrupolar correlation on the sky.  
PTAs are sensitive to gravitational waves with periods ranging from  $1$~month to $20$~yr
 (frequencies between $\approx 2$  and $400$~nHz), with the band constrained by the observing cadence and total timing-campaign duration.
    The strongest gravitational-wave signal in the PTA observing band is predicted to be a stochastic background associated with merging massive black holes \cite[][]{2003ApJ...583..616J,2010arXiv1001.3161S,2012ApJ...761...84R}.
   
   This background imparts a red-noise signal in the TOAs\footnote{In power spectrum analysis, a red noise signal  has more power at lower fluctuation frequencies and is therefore strongly correlated between observations.}.
   This is in contrast with  white signals that are uncorrelated between observations. 
The root mean squared (rms)  amplitude of the background in the residual TOAs is predicted to be $\lesssim 20$~ns over a $5$~year observing span (\citealt{sc2010}, henceforth referred to as Paper I).
  It is currently expected that $20$~to~$100$ MSPs, with timing precision between $\approx 10$ and $100$~ns level are required  
 to significantly detect  gravitational radiation (\citealt[][]{2005ApJ...625L.123J}, Paper I, \citealt[][]{2012ApJ...750...89C}, \citealt{2013CQGra..30v4015S}).
   Constraints on the amplitude of the background  indicate that PTAs are sensitive to GWBs at astrophysically plausible levels \cite[][]{2013Sci...342..334S}   and are already in tension with some models of Galaxy-black hole coevolution %
\cite[][]{2015Sci...349..1522S, 2016ApJ...821...13A}

 To reach the required precision for detecting GWs, it is necessary to identify and correct for many perturbations to the TOAs. 
 These perturbations are incorporated into sophisticated algorithms that incorporate maximum-likelihood or Bayesian approaches. 
Some of these perturbations are deterministic in time, independent of observing frequency, and hence can be parameterized and directly modelled \cite[][]{2006MNRAS.372.1549E}  or  marginalized analytically \cite[][]{2014MNRAS.437.3004L}.  
For example, the perturbation  associated with the secular spin down of the pulsar is modelled through the inclusion of a quadratic polynomial in the timing model. 
Other perturbations exist that are stochastic in time but have a known frequency dependence and can be corrected or constrained using multi-frequency observations.  
These are associated with refraction and diffraction of radio waves in the ionised interstellar (ISM), interplanetary, and ionopsheric media. 
 In this paper, we focus on understanding propagation delays associated with the ISM.

Uncorrected perturbations contribute an additional error to TOAs and degrade the sensitivity of a PTA to GWs.   
It is of particular interest  to identify and correct red-noise perturbations because these more severely  affect the sensitivity of a PTA to a stochastic gravitational wave background than white noise (Paper I).
  
  As part of the effort to detect GWs with pulsars, we are assessing stochastic perturbations to pulsar TOAs in PTA observations.
 In Paper I, we estimated the levels of intrinsic spin noise in MSPs.
 Based on  the levels of timing noise in the more slowly spinning and rotationally unstable canonical pulsars, and two MSPs that exhibit timing noise, 
 we concluded that for many MSPs, spin noise is present at levels comparable to the GWB with similar temporal variability. 

\begin{figure}
\begin{center}
\includegraphics[scale=0.4]{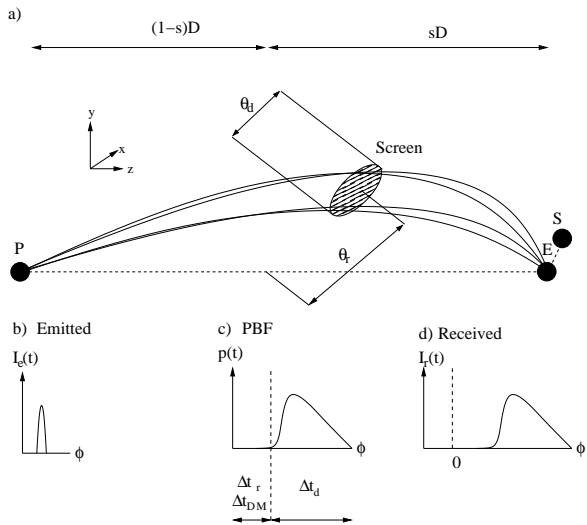} \\ 
\caption{ \label{fig:geometry} 
{\em Panel a:}  Propagation delay geometry.  Observed TOAs are corrected to the solar system barycentre (circle S), assuming that the  radio waves propagate along a direct LOS from the pulsar (circle P) to the Earth (circle E).    The coordinate system orientation is labelled in the upper left corner of the panel. 
  In this analysis,  scattering is assumed to be constrained to a thin screen located a distance $sD$ from the Earth along the pulsar earth LOS, where $D$ is the pulsar-Earth distance.
  At the screen distance, the centre of the tube is offset from the direct LOS by an angle $\theta_r$ and has a width $\theta_d$.  
    The location of the tube, projected on the screen can also be characterised by an image intensity $B(\bm{\theta})$.
  {\em Panels b-d:}  Schematic diagrams of the emitted pulse shape ({\em Panel b}), pulse broadening function $p(\tau,\nu)$ ({\em Panel c}), and observed pulse shape ({\em Panel d}), as a function of pulse phase $\phi = t/P$, where t is the residual time and $P$ is the pulse period.
} 
\end{center}
\end{figure}

 In \cite{cs2010}  (henceforth referred to as Paper II) we  comprehensively assessed the stochastic perturbations to TOAs employing a physical model for TOAs.    
One set of perturbations  discussed in Paper II is associated with the propagation of radio waves through the ISM. 
 As the  radio emission travels from the pulsar to the Earth, it is refracted by the warm ionised electrons (as shown in Figure \ref{fig:geometry}), 
  and the pulse TOA is retarded relative to the expected TOA in vacuum.  
 This delay varies with time because the  sampled region of the ISM changes as the pulsar-Earth line of sight (LOS)  changes due to  relative motion of the Earth and the pulsar through the Galaxy.

The effects of interstellar propagation can be partially mitigated though identification and removal of radio frequency dependent perturbations to the TOAs, because the strength of refraction and hence the magnitude of the TOA perturbation is strongly chromatic.
   This is in contrast with other astrophysically interesting phenomena, such as gravitational radiation, that impart achromatic perturbations.

 
 Multi-frequency mitigation methods are at present only minimally applied in precision-timing observations.
It is usually assumed that chromatic variability in pulse TOAs is associated entirely with the change in the total electron content (the dispersion measure, DM) of the LOS through the interstellar medium\footnote{The interplanetary medium and the ionosphere also contribute to the DM.  However  the contributions are small, do not significantly refract and diffract pulsar radiation at wavelengths of interest, and are therefore not considered here.}, 
and the TOA is  proportional to the DM and the inverse square of observing frequency.  
 When the DM is larger, the pulses arrive slightly later, and vice-versa.  
 This effect is usually assumed to be completely deterministic in frequency; under this assumption TOAs can be corrected by observing at only two frequencies. 
  For the nearby MSPs currently incorporated in PTAs observed at typical observing frequencies,  DM variation delays have an rms amplitude of many microseconds.
Indeed  correcting for DM variations has been identified  as crucial to the success of PTAs \cite[][]{2007MNRAS.378..493Y,2013ApJ...762...94D,2013MNRAS.429.2161K,2014MNRAS.441.2831L}.

The approximation that the ISM is smooth is poor. 
It is well known that the  electron plasma density fluctuations in the ISM cover a wide variety of length scales
\cite[][]{1995ApJ...443..209A}  and that 
they cause multi-path propagation of the pulsar signal \cite[][]{1968Natur.218..920S,1969Natur.221..158R,1990ARAA..28..561R}.  
Thus, in addition to the DM delay, scattering effects must also be considered (\citealt{1984Natur.307..527A}, \citealt{1990ApJ...364..123F}; \citealt{1991ApJ...366L..33H}; \citealt{2010arXiv1005.4914C}, henceforth referred to as C10; Paper II).


In this paper, we  examine some of the  propagation effects associated with diffraction and refraction in the interstellar medium.
We extend on previous studies  of refractive propagation effects.
\cite{1990ApJ...364..123F} investigated propagation effects using a one-dimensional screen. 
They suggested that there were likely significant perturbations to the pulse TOA associated with geometric path length variations that would not be corrected using the simple DM-correction technique.
\cite{1991ApJ...366L..33H} extended this analysis to a two-dimensional screen.
 Both of  these studies were conducted when there were only a few known MSPs and timing precision was much poorer.  
A re-examination of propagation effects is warranted because  of the increase in the number of known MSPs and the improvement in timing precision. 
\cite{2016ApJ...817...16C}
investigated the mis-estimation of TOAs due to incorrect DM estimates.
They modelled the frequency-dependent screen-averaged DM 
and found that when RF bandwidths exceeded an octave in frequency range,  $ \gtrsim 100$~ns timing errors are introduced.
In a more restrictive model, \cite{2015ApJ...801..130L} investigated the effects of non-contemporaneous multi-frequency observations on correcting for DM variations, and found that for widely separated observing epochs, the imprecision of DM correction would limit timing precision for the best pulsar at the levels required to detection gravitational waves.

We quantitatively assess the efficacy of several mitigation strategies relevant to PTAs and other long-term timing observations. 
This study is complementary to the  study  presented in C10, which focused on diffractive effects that cause variations in TOAs on short time scales. 
Due to computational limitations associated with fully diffractive simulations (discussed below) C10   only discussed wave propagation in a narrow frequency band.  
Here we take a complementary approach and focus on effects associated with time and size scales comparable to and larger than  refractive  scales, enabling us to study the effects of large-scale variations  over a wide range of frequencies ($>15:1$ bandwidth).

Our findings are  presented as followed:  
In Section \ref{sec:analytic_delays}, we discuss the relationship between the pulse perturbation and the  image brightness that is the basis of our analysis.   
In Section \ref{sec:refractive_screens}, a refractive screen model for wave propagation is motivated and an approximation for the image intensity  is presented.
In Section \ref{sec:ref_pert}, the TOA perturbations from the  refractive screen are discussed.
In  Section \ref{sec:simuls}, a series of simulations conducted to investigate refractive perturbations are presented.    The details of the simulations are presented in the Appendix.
In Section \ref{sec:mitigate}, mitigation strategies are described.
In Section \ref{sec:future_studies}, we discuss future observations that can confirm the presence of these effects and simulations that guide their analysis.
In Section \ref{sec:conclusions}, we summarise our findings and motivate future studies that complement this work. 

Throughout this paper we use {\em frequency}  in two distinct ways.
 Observing radio frequency  (RF) is denoted as $\nu$.
 Fluctuation frequency, which is the conjugate variable to time in the spectral analyses of  a time series, is denoted as $f$.

\section{Propagation delays \& image intensity}\label{sec:analytic_delays}

 Interstellar propagation contributes both a bulk delay that retards  pulse TOAs and changes the  shape of the received pulse.  
 The effects of interstellar propagation are schematically displayed in Figure \ref{fig:geometry}.
  In panel {\em a}, the observed topocentric TOAs  are measured from  pulsed radio waves that have traveled along a path from the pulsar (circle P) to the earth (circle E).  
  These TOAs are then referred to the the solar system barycentre (circle S).
  Inhomogeneities in the warm interstellar plasma cause the radio waves to travel along an ensemble of paths  depicted as the banana-shaped tube demarcated  by thin solid lines  in Figure \ref{fig:geometry}. 
  In this analysis,  scattering is assumed to be constrained to a thin screen located a distance $sD$ from the Earth along the pulsar earth LOS, where $D$ is the pulsar distance. 
  This geometry is motivated in Section \ref{sec:refractive_screens}.
  
  At the screen, the centre of the tube has a width $\theta_d$ and is offset from the direct LOS  by an angle $\theta_r$;  
  unlike the configuration displayed here, we expect in typical situations $\theta_d > \theta_r$.      
    The shape and location of the tube can be characterised by  its image intensity $B(\bm{\theta})$.

 Figure \ref{fig:geometry}, panels {\em b}-{\em d}  show how the  emitted pulse shape (panel {\em b}) is  affected by the pulse broadening function $p(\tau,\nu)$, ({\em panel c}), resulting in an altered observed pulse shape (panel {\em  d}). 
  Interstellar dispersion and the bulk offset  from the direct LOS delay the TOA of the pulse by amounts labeled $\Delta t_{\rm DM}$ and $\Delta t_r$, respectively.   The width of the tube $\theta_d$ causes the pulse to be broadened, resulting in a further delay $\Delta t_d$.    
  The total delay of the pulse  $\Delta t_{\rm tot}$ is the sum of $\Delta t_d$, $\Delta t_r$, and $\Delta t_{\rm DM}$.
 
The effects of interstellar broadening are expressed using a  pulse broadening function (PBF) $p(\tau,\nu)$, which is also referred to as the impulse-response function.
Using the PBF, the received pulse $I_r(t,\nu)$ at an observation frequency  $\nu$ and time $t$ is related to the emitted pulse $I_e(t,\nu)$ by a convolution
\be
I_r(t,\nu)  = \int d\tau I_e(\tau,\nu) p(t-\tau, \nu).
\ee

The PBF itself is related to the scattered image intensity $B(\bm{\theta})$  by \cite[][]{2001ApJ...549..997C,2006ChJAS...6b.197R}
\be
\label{eqn:pbf_image}
p(t,\nu) = \frac{\int d{\bm \theta} B(\bm{\theta},\nu) \delta(\Delta t_{\bm{\theta}}(\bm{\theta},\nu)  -t)}{\int d{\bm \theta} B(\bm{\theta},\nu)},
\ee
where $\Delta t_{\bm{\theta}}(\bm{\theta},\nu)$ is the net delay through point $\bm{\theta}$ in the image, and $\delta$ is the Dirac delta function.   
We also note that this expression is valid for all scattering strengths and is merely a statement of conservation of energy,  i.e, the flux in the observed pulse is equal to the flux in its image.

Because  the shape of the tube varies with frequency and time, both the structure of the PBF and the shape of the pulse  change.
To determine the effect of the PBF on pulse TOA, it is in general necessary to apply a TOA estimation algorithm, in which the observed pulse is compared to a template \cite[e.g.,][]{1992RSPTA.341..117T}. 
However, if the PBF is much narrower than the observed pulse, the TOA perturbation $\Delta t$ is the mean of the PBF
(\citealt{2008ApJ...674L..37H}; Paper II)
\be
\label{eqn:mean_tau}
\Delta t(\nu)  = \int_0^{\infty} t p(t,\nu) dt. 
\ee

This limit is applicable to current pulsar timing array observations because, for most pulsars currently observed in PTAs, scattering is relatively weak at the typically employed observing frequencies\footnote{We note there have been observations of  large delays from interstellar scattering for some pulsars with lines of sight comparable to MSPs  \cite[e.g.,][]{2010ApJ...708..232B} for which this limit may not be applicable.   A detailed examination of this case is found in paper II.} \
($\nu \approx 0.5$~GHz to $3$~GHz)  and the width of the PBF  is $\ll 1~\mu$s, which is much less than the width of the most narrow ($\approx 50~\mu$s) features in pulsar profiles.

By substituting equation (\ref{eqn:pbf_image}) into equation (\ref{eqn:mean_tau}), the TOA perturbation is 
\be
\label{eqn:delay_intensity}
\Delta t(\nu) = \frac{\int d \bm{\theta} B(\bm{\theta}, \nu) \Delta t_{\bm{\theta}}(\bm{\theta}, \nu)}{\int d \bm{\theta} B(\bm{\theta}, \nu)}.
\ee


In our analysis, equation (\ref{eqn:delay_intensity}) is used to partition the propagation delay into different components,  enabling the identification of unique scaling relationships in both observing frequency $\nu$ and fluctuation frequency $f$. 
Through a  point on the screen $\bm{\theta}$, a  sub-perturbation $i$ contributes a net delay $\Delta t_{\bm{\theta}, i}(\bm{\theta})$.   
The image-averaged perturbation is
\be
\label{eqn:pert_component}
\Delta t_i(\nu) = \frac{ \int d \bm{\theta} \Delta t_{\bm{\theta}, i} (\bm{\theta},\nu) B(\bm{\theta},\nu) }{\int d \bm{\theta} B(\bm{\theta},\nu)}.
\ee

\section{Wave propagation \& thin-screen scattering }\label{sec:refractive_screens}

 The observed emission is instantaneously the combination of scattered radio waves that have traveled in many trajectories through a curved ellipsoidal tube. 
 The tube is characterised by an angular width $\theta_d$, which can be related to the refractive length scale $\ell_r$ on the screen by  $\theta_d \approx \ell_r/(s D)$, which is displayed in Figure \ref{fig:geometry}.


Interstellar propagation effects can be divided into two classes distinguished by their spatial scales.  
Propagation delays on small length scales  (comparable to the width of the tube) result
from  self interference of the radio emission and are  associated with path length differences and electron-density variations along different trajectories within the tube. 
The emission shows variations in frequency because  of the frequency dependence of the phase of the electromagnetic waves and in time because of the changing pulsar-Earth LOS.
  These short term intensity variations are referred to as diffractive interstellar scintillation (DISS).

In addition to causing intensity variations, DISS results in  TOAs variations.
\cite{1990ApJ...349..245C}  examined short-term variations in the TOAs of the MSP B1937$+$21, and showed that  DISS was limiting timing precision at low frequencies. 
C10 examined these effects in simulation and observations and in both cases found a correlation between pulse arrival time and flux for another MSP, J0437$-$4715.  
However  the TOA perturbation was found to be an order of magnitude larger than expected from this relatively weakly scattered pulsar.
\cite{2008ApJ...674L..37H} monitored the slower spinning (canonical)  pulsar B1737$+$13 over half a year and predicted that scattering variations would limit the achieved timing precision if it could be timed to the precision obtained for MSPs. 
  Between observation epochs these diffractive effects are uncorrelated and can be treated as a white noise source.  
 
 On scales larger than $\theta_d$, electron-density variations alter the shape of the tube and its position relative to the direct LOS,  resulting in both variations in flux and the statistics of the DISS. 
  These longer term variations are referred to as refractive interstellar scintillation (RISS).      
 Because these effects are caused by electron-density fluctuations on scales much larger than the diffractive effects,  
  they can introduce red-noise or non-stationary effects into data sets. 

   
\subsection{Modelling  image intensity}


It is necessary in general to model the observed radio emission by integrating over a large three dimensional space \cite[][]{1999ApJ...517..299L}, over the wide range of radio frequencies of interest and long time spans of pulsar-timing campaigns.
Furthermore, because the ISM is turbulent, modelling radio wave propagation beyond second-moment statistical properties necessarily requires simulation.
It is presently computationally prohibitive to conduct a full three-dimensional simulation of wave propagation because of the dynamic range required to model sub-au  scale diffractive variations  to $\sim$kpc distant pulsars. 

To conduct a full simulation requires modelling of wave propagation from length scales as small as the diffractive length scale $\ell_d$, which is the length scale over which  the ISM changes the phase of the radio waves by $\approx \pi$ radians.  
For the lines of sight and observing frequencies relevant to pulsar timing, 
$\ell_d \approx 10^{9}$~cm.  
The largest length  scales are different in each dimension of the simulation.  
Along the LOS (the $z$-axis of Figure \ref{fig:geometry}; the coordinate system is labelled on the top-left corner of panel {\em a}),  the length scale is the distance to the pulsar (nominally $\sim 10^{21}$~cm).   
 In the direction of projected motion (the $x$-axis of Figure \ref{fig:geometry}), 
 the outer scale of the simulation is the projected motion of the pulsar,  ($V_{\rm eff} T \approx 10^{14}$~cm, for $V_{\rm eff}= 100$~km s$^{-1}$, and $T =10$~yr).   
Perpendicular to the projected direction of motion (the $y$-axis of Figure \ref{fig:geometry}), 
the refractive length scale $\ell_r$  (nominally $10^{13}$~cm for the pulsars and observing frequencies relevant here).  
For these typical values, to possess the required spatial dynamic range a computationally impractical  phase screen comprising $10^{22}$ spatial samples is needed.

Two physically valid approximations make the simulations computationally viable.
Firstly, scattering is assumed to be located predominantly in a geometrically thin phase-changing screen, which appears to be the case for some  nearby pulsars \cite[][]{2001ApJ...549L..97S, 2006ApJ...637..346C,2006ChJAS...6b.233P}.  

Even with this assumption, it is unfeasible to conduct simulations over the wide range of frequencies (a factor of~$2$~or greater) necessary for pulsar timing observations over the observing spans of $T=5$~to~$10$~yr relevant for GW detection.     
A computationally viable simulation of a two dimensional screen can be conducted by either reducing the total observing time and range of frequencies modelled or by neglecting diffractive effects.  
C10 chose to reduce both the range of frequencies modelled and total observing span.   
Over a narrow range of frequencies and an observing span of a few months, C10 conducted a full diffractive simulation. 
They additionally assumed that   dispersive delays could be perfectly corrected.   

In the simulations presented here, 
the screen resolution is reduced by modelling only phase variations comparable to or larger than the  refractive length scale $\ell_r$ and using a renormalization method first presented by \cite{1986ApJ...310..737C}.
Because $\ell_r$   is much larger than the diffractive scale,  we can simulate phase screens over a wide range of observing frequencies over $T=10$~yr observing spans, enabling a proper assessment of the non-stationarity of refractive propagation delays.




In Appendix \ref{sec:app_image_matrices}, we present a model for the image intensity $B(\bm{\theta})$ that we summarise briefly here.  
First, the locations of unique sub-images are identified at stationary phase points (SPPs), where the gradient of the total phase (the combined geometric and refractive phase) is zero.  
 These SPPs identify regions  where the phase is slowly changing  and hence the waves constructively interfere. 
Around each SPP, the refractive phase is modelled using a paraboloid.
It is assumed that the unmodeled diffractive phase variations follow ensemble-average behaviour, with a non-refracted image intensity $B_0(\bm{\theta})$.  
Under these assumptions, the wave propagation through the screen can be calculated analytically using the Kirchhoff diffraction integral (KDI), and the intensity of the refracted image $B(\bm{\theta})$ expressed as a function of the non-refracted image. 
With the paraboloidal approximation, there are two modifications to the non-refracted image.  
  Firstly,  relative to the direct line of sight, the image centre is offset to the stationary phase point $\bm{\theta_0} = \bm{r_0}/(s D)$. 
 Secondly,  relative to the unperturbed image, the image is rotated and stretched along two principal axes, parametrized by a $2\times2$ matrix $\bm{M}$.
With this approximation, the refracted image can be expressed as a function of the non-refracted image,  
\be
\label{eqn:refracted_image}
B(\bm{\theta},\nu) = B_0\left(\bm{M}[\bm{\theta} - \bm{\theta_0}],\nu \right).
\ee

The  image distortion  also causes flux-density variability that may be correlated with the TOA perturbations. 
Relative to the non-refracted image, the flux density is
\be
F_{\rm rel}(\nu) = \frac{\int d\bm{\theta} B(\bm{\theta},\nu)}{\int d\bm{\theta} B_0(\bm{\theta},\nu)}.
\ee





With this model for image intensity, we can calculate the delays through the screen.
As these expressions  model only parabolic variations in refractive phase, they underestimate the strength of effects near caustics and other situations in which diffraction plays a strong role. 
To study these effects it is necessary to employ a fully diffractive simulation, which is beyond the scope of this paper. 


\section{TOA perturbations}\label{sec:ref_pert}

\subsection{Dispersive perturbations}

Through a point $\bm{\theta}$ on the screen the TOA perturbation associated with radio-wave dispersion is
\be
\Delta t_{{\rm DM}, \bm{\theta}}(\nu) = -\frac{2 \pi}{\nu} \phi_r(\bm{\theta},\nu),
\ee
where $\phi_r$ is the dispersive contribution to the screen phase.  
This perturbation is anti-correlated with the screen phase because the delay is associated with the group velocity (and not the phase velocity) of the propagating radio waves.

Averaged over the entire image, the delay is
\be
\label{eqn:disp_delay}
\Delta t_{\rm DM}(\nu) =  -\left(\frac{ 2\pi }{\nu}\right) \frac{\int  d \bm{\theta} B(\bm{\theta},\nu) \Phi_{r} (\bm{\theta})}{\int  d \bm{\theta} B(\bm{\theta},\nu) }.
\ee

This  delay is both deterministically and stochastically wavelength dependent.  
By expanding about the direct LOS, the different contributions are identified:
\be
\label{eqn:disp_delay_exp}
\Delta t_{\rm DM}&&= -\left(\frac{ 2\pi }{\nu}\right)  \Phi(0)  - \left(\frac{ 2\pi }{\nu}\right) \Phi_{\rm SPP}(0) \nonumber \\
&&~~~-\left(\frac{ 2\pi }{\nu}\right) \frac{\int d \bm{\theta} B(\bm{\theta}) \left[ \Phi(\bm{\theta}) -\Phi_{\rm SPP}(0)-\Phi(0)\right] }{ \int d \bm{\theta} B(\bm{\theta})} \nonumber\\
&&= \Delta t_{\rm DM} + \Delta t_{{\rm DM}, C} + \Delta t_{{\rm DM}, I}.
\ee   
The first term  in equation (\ref{eqn:disp_delay_exp}),  $\Delta t_{\rm DM}$,  is the group delay along the direct LOS.  
It is deterministically  correlated between observations at different frequencies, and is completely removed by the standard DM correction scheme that uses TOAs at two or more widely separated frequencies to estimate DM  \cite[][]{2007MNRAS.378..493Y,2013MNRAS.429.2161K}.    
The second term, $\Delta t_{{\rm DM}, C}$, is the delay through the centre of the image (i.e, the delay through the SPP).  
This term is not perfectly correlated between different frequencies because the location of the SPPs vary  with  frequency.
The third term,   $\Delta t_{{\rm DM}, I}$, is associated with the averaging of the phase across  the image $B(\bm{\theta})$, and also  changes with frequency and time.  
At lower frequencies the scattered image is larger,  and hence the delay is averaged over a larger region of the phase screen. 
Additionally, the shape and the location of the image decorrelate with increasingly widely separated frequencies 
\cite[][]{2016ApJ...817...16C}.

\subsection{Geometric delays}

There are TOA variations associated with the variable path length through the refracted image that we classify as geometric perturbations.
Pulse TOAs are determined at the observatory and then referred to the solar system barycentre (SSB) using a model for the position of the Earth.  
This correction depends on the assumed position of the pulsar, which, for MSPs, is typically found to highest precision by modelling the TOAs.

The perturbations associated with refraction are  the difference between the propagation time in the presence and absence of the screen. 
In the absence of the screen, the radio waves travel directly from the pulsar to the Earth. 
We define a coordinate system with the SSB at the origin, 
the pulsar at  position ${\bm D} = D \bm{\hat{n}}$, 
and the Earth is at a time-variable location $\bm{r}_{\earth}(t)$.      
 For clarity of presentation  the pulsar is assumed to be stationary, which does not affect our analysis.  
 
In the absence of a refracting screen, the propagation time from the pulsar to the solar-system barycentre is
\be
\Delta t_{\sun} = \frac{D}{c}
\ee
and the propagation time from the pulsar to the Earth is
\be
\label{eqn:earth_noscatter}
\Delta t_{\earth} &&= \frac{1}{c} \left| {\bm D} -{\bm r_{\earth}}  \right| \nonumber \\
&&\approx \frac{D}{c}\left[ 1+\frac{1}{2}\left( \frac{r_{\earth}^2}{D^2} - \frac{2 {\bm r_{\earth}}\cdot \bm{\hat{n}}}{D} \right)\right].
\ee
Therefore in the absence of the screen, a correction $\Delta t= \Delta t_{\earth} -\Delta t_{\sun}$ is applied to refer the arrival times to the SSB:
\be
\label{eqn:toa_noscatter}
\Delta t = -\frac{1}{c}\left( {\bm r_{\earth}}\cdot {\bm \hat{n}}\right)+ \frac{1}{2c} \frac{r_{\earth}^2}{D}.
\ee
The first term  is associated with the pulsar's sky position and results in an $1$~yr periodic variation in the TOAs.
The second term is constant  assuming  the Earth is in a circular orbit.  

In the presence of a refracting screen, the path length to the Earth is increased relative to the direct LOS.  
The travel time through a point in the screen offset by $\bm{r}$  from the direct LOS is
\be
\label{eqn:earth_scattering}
&&\Delta t_{\earth, s}(\bm{r}) =  \frac{1}{c} \left | (1-s)D \bm{\hat{n}} + \bm{r}  \right| + \left| s D \bm{\hat{n}}  + \bm{r} - \bm{ r_{\earth}} \right| \\
&&\approx \frac{D}{c}  +\frac{1}{2c} \left[ 2 (\bm{r}-\bm{r_{\earth}})\cdot\bm{\hat{n}}  + \frac{r^2}{(1-s)D} + \frac{(\bm{r}-\bm{ r_{\earth}})^2}{s D} \right]. \nonumber
\ee

The total geometric perturbation is the difference between equations (\ref{eqn:earth_scattering}) and (\ref{eqn:earth_noscatter}):
\be
\Delta t_g(\bm{r}) &=& \frac{1}{2 c} \Biggl[ 2 {\bm r} \cdot \bm{\hat{n}} + \frac{r^2 - 2(1-s) {\bm r}\cdot{\bm r_{\earth}}}{s(1-s)D}  \nonumber    \\
&& ~~~~~~~~+ \left( \frac{1-s}{s}\right) \frac{r_{\earth}^2}{D} \Biggr]. 
\ee

Substituting $\bm{\theta} = \bm{r}/(sD)$, and noting that $\bm{r} \cdot \bm{\hat{n}} = 0$, the total geometric perturbation is 
\be
\label{eqn:dt_angle}
\Delta t_{\bm{\theta},g} ({\bm \theta}) = \left(\frac{D}{2 c}\right) \left(\frac{s}{1-s}\right) \theta^2 - \frac{1}{c} {\bm \theta} \cdot {\bm r_{\earth}}  + \left(\frac{1-s}{2s}\right) \frac{r_{\earth}^2}{c D}. 
\ee
The first term in equation (\ref{eqn:dt_angle}) is the geometric delay through the screen.
The second term is the result of incorrectly referencing the arrival time of the pulse at the solar system barycentre.
The final term  is nearly constant in time  because the Earth's motion about the SSB is nearly circular and is henceforth ignored.  

In the case of a single stationary phase point (i.e., a single sub-image), the geometric delay averaged over the image is found by substituting equation (\ref{eqn:dt_angle}) into equation (\ref{eqn:pert_component}) and integrating over $\bm{\theta}$,
\be
\label{eqn:geo_delay}
\Delta t_g (\nu) &=& \left(\frac{D}{2c} \right) \left( \frac{s}{1-s}\right) \theta_r^2  - \frac{1}{c} \bm{\theta_r} \cdot \bm{r_{\earth}} \nonumber \\
&&~~+ \frac{1}{\det \bm{M}}  \left(\frac{D}{2c} \right) \left( \frac{s}{1-s}\right) \theta_d^2,
\ee
where it is assumed that the non-refracted image is centred on the origin, i.e.,  
\be
\int  \bm{\theta} B_0(\bm{\theta})d\bm{\theta} =0,  
\ee
$\bm{\theta_r}$ is the centre of the refracted image (as expressed  in equation \ref{eqn:spp_centre} in the Appendix),  and $\theta_d^2$ is the mean squared deflection angle,
\be
\theta_d^2 \equiv \frac{\int \theta^2 B_0(\bm{\theta}) d\bm{\theta}}{\int  B_0(\bm{\theta}) d\bm{\theta}}. 
\ee
We emphasise that $\bm{\theta_r}$, $\theta_d$, and $\bm{M}$ all vary with observing frequency and epoch, and that the perturbations will contain terms that are at least partially stochastic in frequency.
 

\subsection{Total delay and scaling relationships}

The  total delay $\Delta t_{\rm tot}$  is the sum of  the perturbations associated with dispersion found in Equation (\ref{eqn:disp_delay}), and the geometric and diffractive delay terms  found in Equation (\ref{eqn:geo_delay}).     
These perturbations have fluctuations that vary with both radio frequency and time.
In Table \ref{tab:scaling} we summarise the scaling laws in radio frequency and fluctuation frequency. 
We show the scalings for the image shape parameters: phase $\Phi$, the components of the vector $\bm{\theta_0}$, and matrix $\bm{M}$.   We also show the scalings for the TOA perturbations which are in general functions of these terms.  
To estimate the scalings, we assume that the phase  has a power-law power spectrum (equivalent to assuming that electron-density variations also have power-law statistics) and that its structure function can be expressed as  
\be
D_\Phi(\bm{b}) \equiv \langle \Phi(\bm{r})\Phi(\bm{r} + \bm{b}) \rangle  = \pi^2  \left(\frac{b}{\ell_d}\right)^{\beta-2},
\ee
where $\ell_d$ is the diffractive scale.   The exponent $\beta$ is expected to range between $3$ and $4$, with $\beta=11/3$ in a Kolmogorov medium.
  
  The perturbations show three distinct scalings with radio frequency, suggesting that observations at minimally four frequencies are potentially necessary to recover an achromatic signal of interest within the TOAs.
     Because some of the scalings depend on the power spectrum of the phase variations, proper correction requires that ancillary observations be used to measure $\beta$, or that  $\beta$ is included as a parameter in the TOA modelling.  This can accomplished by  using the TOA observations themselves (requiring observation at a fifth observing frequency), or analysis of ancillary information such as DISS. 

 The perturbations show a wide variety of time variability because they have power spectra with spectral indexes nominally between $-2$ and $+2$. 
The perturbations also show non-stationarity that is different  than expected from the gravitational wave background, which has a power spectrum with spectral indices between  $-4$~to~$-5$, for background formed from MBH binaries \cite[][]{2003ApJ...583..616J} and cosmic strings \cite[][]{2005PhRvD..71f3510D}, respectively.  

\begin{table}
\begin{centering}
\caption{Radio and Fluctuation Frequency Scaling Laws}
\label{tab:scaling}
\begin{tabular}{llrr}\hline\hline\\
\multicolumn{1}{c}{Property}  & \multicolumn{1}{c}{Symbol} &\multicolumn{1}{c}{RF} &\multicolumn{1}{c}{FF}  \\ \hline
\multicolumn{4}{c}{Shape}\\
\hline
Phase &  $\Phi$ & $-2$ & $-\beta$ \\
 Refraction Angle & $\bm{\theta_r}~(\propto \partial \Phi)$ & $-2$ &$-\beta+2$ \\
Curvature & $\bm{C}~(\propto \partial^2 \Phi)$ & $-2$ & $-\beta+4$ \\
 Diffraction Angle & $\theta_d$ & $-\gamma$  &$0$ \\
    \hline 
    \multicolumn{4}{c}{TOA Perturbations}  \\
    \hline
DM  through LOS&  $\Delta t_{\rm DM}$ & $-2 $  &$-\beta$   \\
DM through SPP &   $\Delta t_{\rm DM,C}$ & $-2 $  &$-\beta+2$   \\
DM  through screen &       $\Delta t_{\rm DM,S}$ & $-2 -\gamma $  &$-2\beta+4$   \\
Barycentering Error & $\Delta t_{\rm Bary}$ &  $-2 $  & $-\beta+2$  \\
Geometric Delay & $\Delta t_{\rm Geo}$ & $-4$  &$-2\beta+4$   \\
PBF variations & $\Delta t_{\rm Diff}$ & $-2 -\gamma$  &$ -\beta+4$ \\
 \hline
\multicolumn{4}{c}{Gravitational Wave Background}\\
\hline
 GWB  & \nodata & 0 & $-4$ to $-5$\\ \hline
\end{tabular}
\end{centering}
\begin{flushleft}
 Scaling laws for the strengths of different perturbations in radio frequency (RF)  and fluctuation frequency (FF).   The medium is assumed to have  density fluctuations with a power-law  wavenumber spectrum $P_\Phi \propto q^{-\beta}$. 
  In the top row, the scaling with respect to properties of the refracted image are displayed.   
 In the bottom rows, the scalings for the various TOA perturbations are displayed.   Note that $\gamma \equiv 2\beta/(\beta-2)$.  
 \end{flushleft}
  \end{table}

\section{Simulating refractive perturbations}\label{sec:simuls}

We conducted a series of simulations for a wide variety of screen and observing frequencies to determine the strength of the propagation effects and assess the efficacy of mitigation techniques.  
A detailed description of the implementation is presented in  Appendix \ref{sec:simuls_appendix}.
  By construction, the pulsar and SSB are held fixed relative to a moving phase screen that incorporates actual motions of the pulsar and SSB.  Therefore,  as a function of time the phase is $\Phi(\bm{x},t) = \Phi(\bm{x}-s \bm{V_p} t - (1-s) \bm{r}_{\earth}(t) )$,  where $\bm{r}_{\earth}(t) $ is the Earth's position relative to the solar system barycentre perpendicular to the LOS and $\bm{V_p}$ is the pulsar's transverse velocity. 

\subsection{Generating phase screens}

\begin{figure}
\begin{center}
\includegraphics[scale=0.5]{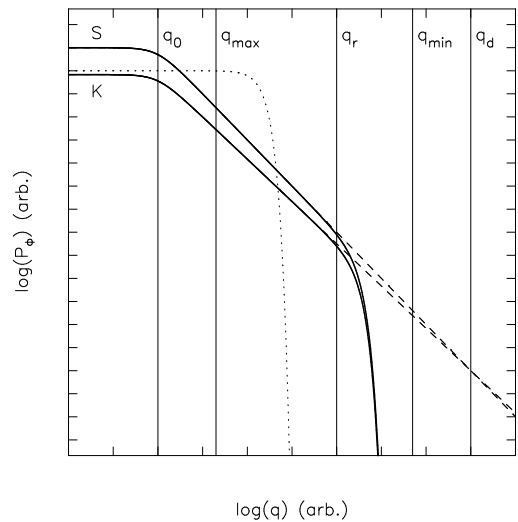} \\ 
\caption{ \label{fig:powspec} 
Schematic power spectrum $P_\phi(q)$ versus spatial frequency $q$ for square-law ($P_\phi \propto q^{-4}$, upper curve {\em S}) and Kolmogorov ($P_\Phi \propto q ^{-11/3}$, lower curve K) media.  The media are calibrated such that they have the same size fluctuations at the diffractive scale $\ell_d = 1/q_d$. 
The simulations only include structures between the reciprocal of the maximum length scale  $q_{\rm max} = 1/L = 1/V_{\rm eff} T$, where $V_{\rm eff}$ is the effective motion of the LOS across the screen and $T$ is the total observing span; and the reciprocal of the minimum length scale in the medium $q_{\rm min}$, which is the refractive scale at the highest observing frequency.  
While the electron density contains fluctuations over a wide range of scales (thin dashed lines), the phase fluctuations are smoothed to the refractive scale to properly model DISS smoothing (thick lines). 
The dotted line represents the power spectrum used for the smooth media simulations, discussed in Section \ref{sec:phen_smooth}.
} 
\end{center}
\end{figure}

Phase variations are modelled on a grid, 
with grid size and spacing constrained by the required spatial dynamic range. 
     The screen must be sufficiently large to model the largest relevant length scales: 
perpendicular to the effective motion, the screen must be much larger than  the image size at the lowest frequency (the frequency at which the image has the largest size).
In the direction of projected motion, the screen must be sufficiently large to model the motion of the pulsar across the screen or the refractive scale at the lowest frequency, whichever is larger.      
The screen must also have sufficient resolution to over-resolve the image at the highest observing frequency 
(the frequency at which the refractive scale is the smallest).
Fluctuations are generated to follow a predetermined ensemble-average fluctuation power spectrum $P_\Phi(q)$.
Additionally, we assume that the outer scale of the medium $L_0 = 1/q_0$ (the length scale at which the phase variations become stationary) is larger than the screen size $L_{\rm max} = 1/q_{\rm max}$ and the inner scale $L_{\rm in}$  is much smaller than the smallest scale of the simulation $L_{\rm min} = 1/q_{\rm min}$.   
To model the smoothing caused by DISS, a Gaussian low-pass filter is applied to the screen in the Fourier domain $W(q)  = \exp( -q^2/2 q_r^2)$ where $q_r = 1/\ell_r$. 
 In Figure \ref{fig:powspec} we display the schematic ensemble-average power spectrum of the fluctuations and relevant lengths scales of the simulation.
 

   
\subsection{Calculating the perturbations}

At each time and observing frequency, the refracted image $B(\bm{\theta})$   is calculated  by locating stationary phase points (SPPs) using a grid search. 
Around each  SPP  the phase $\Phi_r$ is approximated with a paraboloid.
The total perturbation is found by numerically integrating Equations (\ref{eqn:disp_delay}) and (\ref{eqn:geo_delay}) over the scattered sub-images formed around each SPP.
In the simulations presented here, we assume that there is only one SPP per epoch.  

\subsection{Simulation properties}

We investigated a variety of scattering screen and observing frequencies.
The simulations were based on media with fluctuations with  power-law spectra and a range of screen strengths.  
The screen strengths were defined using the  diffractive scintillation bandwidth $\Delta \nu_d$ at a fiducial observing frequency $\nu_{\rm REF}$.

We simulated observing strategies based on those employed in current PTAs \cite[][]{2010CQGra..27h4013H,2013ApJ...762...94D,2013CQGra..30v4009K,2013PASA...30...17M} and those likely to be employed with PTAs on planned facilities such as the Square Kilometre Array
\cite[SKA, ][]{2015aska.confE..37J}.
We simulated perturbations over a broad range of frequencies (enabling us also to investigate the effect of observing bandwidth), and levels of additive white noise. 

For the simulations presented here, TOAs at a  total of $N_{\rm TOA} = 500$ epochs were simulated at evenly spaced intervals over a $T=5$~yr observing span.  
This observing span matches nominal PTA observing campaign  specified for the detection of the stochastic GWB with moderate confidence \cite[paper I, ][]{2005ApJ...625L.123J}, but is shorter than current timing-array campaigns. 



\subsection{Phenomenology}

\subsubsection{Smooth media}\label{sec:phen_smooth}

We first generated smooth screens that only have fluctuations on spatial scales comparable to the screen size. 
Further details of this implementation are provided in Appendix \ref{sec:simuls_appendix}.
We found that in the simulations, the perturbations followed the expected frequency scalings displayed in Table \ref{tab:scaling}, indicating that our code is correctly implementing the model outlined above.

\subsubsection{Turbulent media}\label{sec:phen_turb}

\begin{figure*}
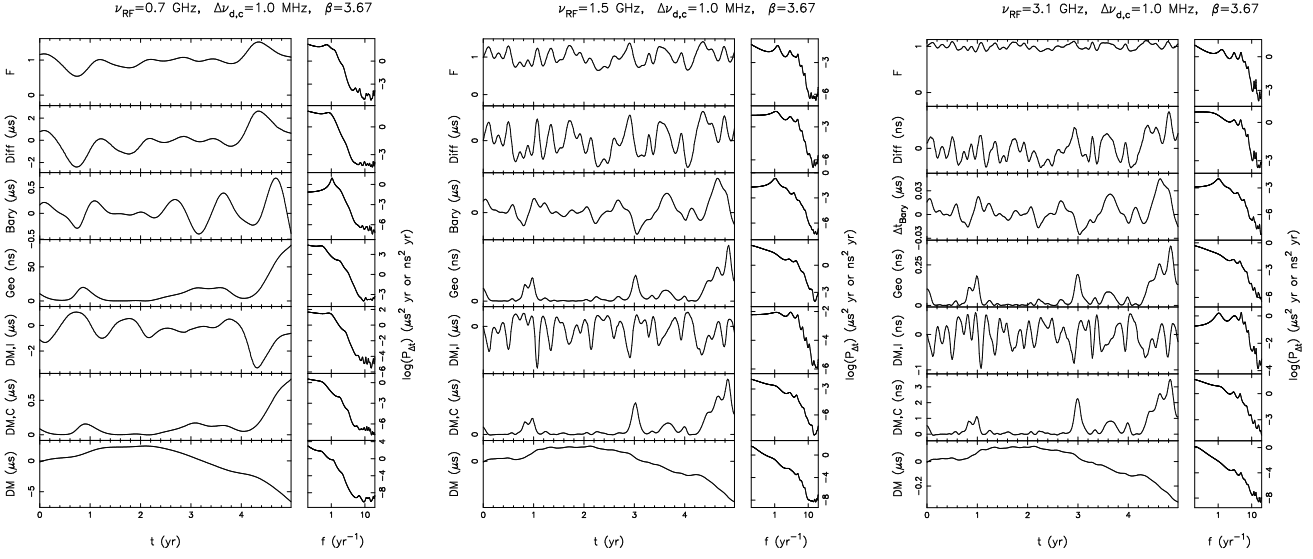

\begin{center} 
\begin{tabular}{ccc}
\includegraphics[scale=0.45]{{1937_0.7_2}.eps} &\includegraphics[scale=0.45]{{1937_1.5_2}.eps}  & \includegraphics[scale=0.45]{{1937_3.1_2}.eps}  \\
\end{tabular}
\caption{ \label{fig:TOA_pert_k}  ISM-induced perturbations to TOAs for a Kolmogorov medium at observing frequencies $\nu_{\rm RF}= 0.7$~GHz (left panel),  $1.5$~GHz (centre), and $3.1$~GHz (right). The level of scattering is similar to that of the MSP B1937+21.  From bottom to top, we display the dispersive perturbation through the direct LOS $\Delta t_{{\rm DM}}$ (DM), the dispersive perturbation associated with the centre of the image $\Delta t_{{\rm DM}, C}$ (DM,C), the dispersive delay associated with image averaging $\Delta t_{{\rm DM},I}$ (DM,I),     the path-length variation associated with the offset of the stationary phase point $\Delta t_{\rm Geo}$ (Geo), the barycentric perturbation term $\Delta t_{\rm Bary}$ (Bary), and the geometric perturbation associated with the averaging image $\Delta t_{\rm Diff}$ (Diff). For the plots of $\Delta t_{\rm Diff}$ we have subtracted the average  $\Delta t_{\rm Diff, avg}$ for clarity.
We also display the relative flux  $F_{\rm rel}$ normalised to the non-diffracted flux (F).  We show maximum-entropy power spectra to the right of each time series.  Both the abscissa and ordinate are displayed with logarithmic scaling.    The amplitudes of the power spectra are labeled in logarithmic units to the right of the plots in either $\mu{\rm s}^2~{\rm yr}$ or ${\rm ns}^{2}~{\rm yr}$. In the case of the power spectra of the flux, the power spectra are in arbitrary units. }
\end{center}
\end{figure*}

\begin{figure*}
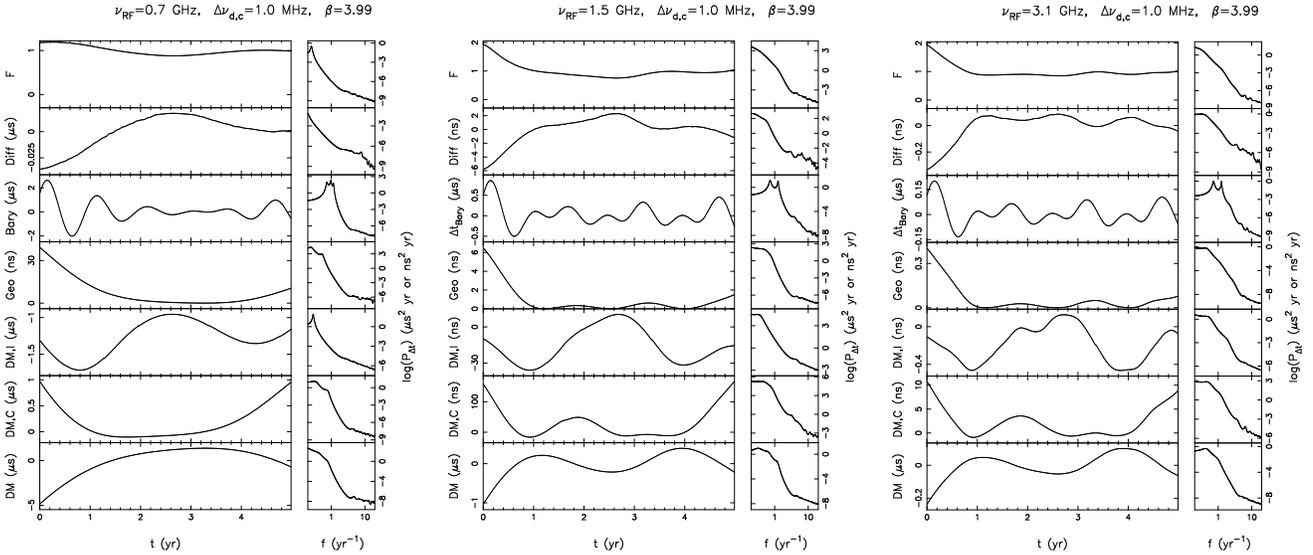

\begin{center} 
\begin{tabular}{ccc}
\includegraphics[scale=0.45]{{1937_square_0.7_2}.eps} &\includegraphics[scale=0.45]{{1937_square_1.5_2}.eps}  & \includegraphics[scale=0.45]{{1937_square_3.1_2}.eps}  \\
\end{tabular}
\caption{ \label{fig:TOA_pert_s}   ISM-induced perturbations to TOAs for a square-law medium at observing frequencies $\nu_{\rm RF}= 0.7$~GHz (left panel) and $1.5$~GHz (centre), and $3.1$~GHz (right panel). The plots are labeled the same as for Figure \ref{fig:TOA_pert_k}.   }  
\end{center}
\end{figure*}

Simulations of  single realisations of TOA perturbations associated with Kolmogorov and square law turbulent media are displayed in Figures \ref{fig:TOA_pert_k} and \ref{fig:TOA_pert_s}, respectively, for scattering screens with the strength expected from PSR~B1937$+$21.
It has the sixth-largest  DM of the pulsars in current PTA samples \cite[][]{2013CQGra..30v4010M} and  is also one of the most scattered pulsars observed with $\Delta \nu_d \approx 1$~MHz at a frequency of $1.4$~GHz \cite[][]{2013MNRAS.429.2161K}.
 While the best pulsars have lower DM and are less scattered than PSR~B1937$+$21, it is  possible that more pulsars with this level of scattering (or greater) will be incorporated into PTAs as more highly dispersed MSPs are discovered. 
 We simulated observations at $12$ frequencies between $0.7$ and $3.1$~GHz, nominally covering the frequency range of current PTA efforts, and a frequency range likely employed in future broad-band observations of the pulsar.
 
\begin{figure*}
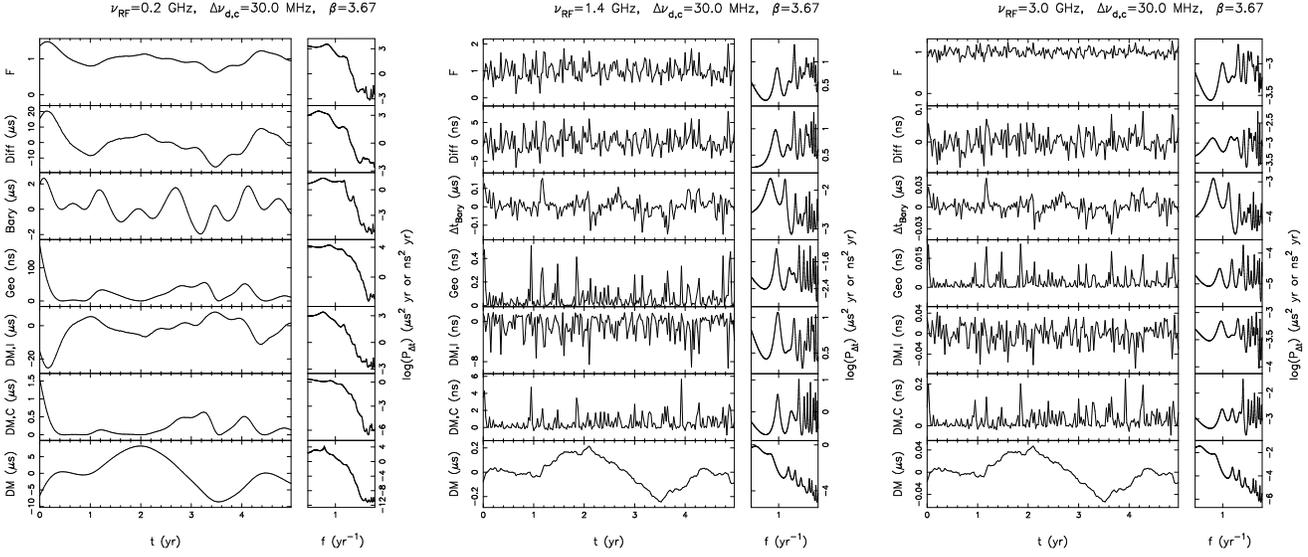

\begin{center} 
\begin{tabular}{ccc}
\includegraphics[scale=0.45]{{1713_0.2_2}.eps} &\includegraphics[scale=0.45]{{1713_1.4_2}.eps}  & \includegraphics[scale=0.45]{{1713_3.0_2}.eps}  \\
\end{tabular}
\caption{  \label{fig:TOA_pert_1713}  ISM-induced perturbations to TOAs for a  Kolmogorov medium, but with PSR~J1713$+$0747-strength scattering  at observing frequencies $\nu_{\rm RF}= 0.2$~GHz (left panel) and $1.4$~GHz (centre), and $3.0$~GHz (right panel). The plots are labeled the same as for Figure \ref{fig:TOA_pert_k}.   }  
\end{center}
\end{figure*}

 In Figure \ref{fig:TOA_pert_1713}, we show perturbations for the more weakly scattered PSR~J1713$+$0747 over a wider range of frequencies. 
 PSR J1713$+$0747  shows the best timing precision in current timing-array efforts 
 \cite[][]{2013ApJ...762...94D, 2015ApJ...813...65T}. 
 It also shows only modest levels of scattering with $\Delta \nu_d \approx 30$~MHz at a frequency of $1.4$~GHz \cite[][]{2013MNRAS.429.2161K}.    
 For this pulsar, we simulated observations at $15$ frequencies between $0.2$ and $3.0$~GHz.    These simulations enable us to assess the utility of observations  with metre-wavelength arrays currently being commissioned and the low-frequency component of the Square Kilometre Array (SKA).

For each realisation, the perturbations are displayed at three frequencies spanning the simulated range. 
Here we highlight a few qualitative properties of the perturbations, but note that large realisation-to-realisation variations in the strength and nature of the perturbations are exhibited because of the stochasticity of the phase screens. 

{\em Strength and stationarity:} The rms values of the different perturbations range over an order of magnitude, depending on the strength of the scattering and the RF.  
Variations in DM  contribute  perturbations that  have both the largest amplitude and the most non-stationary (i.e., most red) behaviour. 
 Most but not all of  the DM variations are correlated between observing bands.  
 To better identify the components of the perturbation that correlated and uncorrelated in observing frequency, we  show the two terms of the dispersive delay identified in equation (\ref{eqn:disp_delay_exp}):  the dispersive delay through the direct LOS $\Delta t_{\rm DM}$, through the image centre $\Delta t_{\rm DM,C}$, and averaged over the image $\Delta t_{\rm DM, I}$.    
The barycentric-correction term $\Delta t_{\rm Bary}$ shows modestly periodic variations if the image remains in the same offset position for many years. 
For pulsars moving slowly relative to the Earth in the presence of screens with strong gradients, the DM variations can show annual variations as the pulsar-earth LOS annually revisits the same electron column density \cite[][]{2013MNRAS.429.2161K}. 

{\em Frequency Scaling:}  TOA variations are larger and smoother at lower frequencies.
The frequency scaling of the rms amplitudes of the perturbations are weaker than predicted in Table \ref{tab:scaling} because diffraction smooths over the scattering screen.

{\em Correlation between $\Delta t_{\rm DM,S}$ and $\Delta t_{\rm Diff}$:}
The dispersive delay associated with the centre of the stationary phase point $\Delta t_{\rm DM,C}$ is directly correlated with the geometric delay, which was also noted in C10 and Paper II.

{\em  Differences between Kolmogorov and square law media:} 
For the same scattering strength at a fixed reference frequency, 
square-law media show smoother perturbations because the screen contains greater  power on larger spatial scales.    
The barycentric correction term has a larger and more periodic component in the  simulation because prominent large-scale structures refract the SPP farther away from the direct LOS for longer periods of time (in some cases many years).    
Annual variations are observed when diffractive smoothing is larger than the  distance the pulsar moves on the screen in the year, so are more apparent at lower frequencies.

{\em Correlation between  $\Delta t_{\rm Diff}$ and Flux Density:}   
There have been previous reports of an anti-correlation between pulse arrival and refractive flux density variations \cite[][]{1998A&A...334.1068L} and  during extreme scattering events \cite[][]{1993Natur.366..320C}. 
In the refractive simulations presented here, there is a modest correlation between flux density (in the uppermost panels of Figure \ref{fig:TOA_pert_s} and \ref{fig:TOA_pert_k}) and $\Delta t$.  
For square-law media there is a large variation in the level of correlation from realisation to realisation. 
 For Kolmogorov simulations there is a $20\%$ to $40\%$ anti-correlation between flux density and dispersive delay $t_{\rm DM}$, consistent with observations of PSR~B1937$+$21 reported in  \cite{1998A&A...334.1068L}.   
 Refractive flux-density variations are difficult to measure in practice because DISS introduces a large estimation error on the refracted flux. 

   We also note the presence of peaks in the power spectra at frequencies near $1$~yr$^{-1}$ that are associated with the effect of the Earth's motion about the solar-system barycentre.

\section{Mitigating refractive perturbations}\label{sec:mitigate}


\subsection{Mitigation methods}

We investigated several  strategies for estimating infinite-frequency TOAs $t_\infty$ using only the measured TOAs. 
At each epoch,  TOAs are measured at discrete frequencies.
A general model for single-epoch TOAs is
\be
t(\nu) = t_\infty + \sum_{i=1}^{N_c} C_i \nu^{-X_i}, 
\ee
where $t_\infty$ is the  estimated infinite-frequency arrival time and the $i=1, N_c$ additional terms  model chromatic TOA variations with unique power-law spectral scalings $X_i$.  
 In the next sections we  explore the appropriate number of terms $N_c$ and spectral indices to include in arrival-time modelling.  
Within the modelled phase screens there are five deterministic frequency scalings: the achromatic term ($t_\infty$)   and terms proportional to $\nu^{-X_i}$, where $X_i=-2, \approx -4.4$, and $-6.4$.

If no chromatic terms are included, the model for the time of arrival $t(\nu)$ at frequency $\nu$ is simply
\be
\label{eqn:mitigate_nocor}
t(\nu) = t_\infty^{(1)},
\ee
where the superscript labels this as model (1). 
This model is applicable when chromatic perturbations are very small relative to other TOA errors such as those due to receiver noise.
  
 In most current precision timing observations, TOAs are corrected assuming that the only chromatic perturbations is proportional to $\nu^{-2}$ which is assumed to be associated with DM variations, though it will also correct for the barycentric error.   In this case model (2) is 
  \be
\label{eqn:mitigate_2freq}
t(\nu) = t_\infty^{(2)} + C_{-2}^{(2)} \nu^{-2}.
\ee

A three-term mitigation strategy includes an additional term that scales  $\propto \nu^{-X_3^{(3)}}$ with a value dependent on the LOS structure:
\be
\label{eqn:mitigate_3freq}
t(\nu) = t_\infty^{(3)} + C_{2}^{(3)} \nu^{-2} + C_{3}^{(3)} \nu^{-X_3^{(3)}}, 
  \ee
In a previous analysis, \cite{1990ApJ...364..123F} used $X_3^{(3)}=4$ because they assumed that the geometric term $\Delta t_{\rm Geo}$ has the largest contribution after the dispersive delay.

 Similarly, the four-term model has two terms in addition to a $\nu^{-2}$ scaling:
\be
\label{eqn:mitigate_4freq}
t(\nu) =t_\infty^{(4)} + C_{2}^{(4)} \nu^{-2} + C_{3}^{(4)}  \nu^{-X_3^{(4)}}   + C_{4}^{(4)} \nu^{-X_4^{(4)}},
\ee
where the frequency scalings of  $X_3^{(4)}$, $X_4^{(4)}$   depend on the properties of medium, because these will determine which particular perturbing term dominates the TOA variations.

We consider two TOA weighting schemes used to conduct the least-squares fits to the models presented in Equations  (\ref{eqn:mitigate_nocor}) to (\ref{eqn:mitigate_4freq}).  In the first, the TOAs are weighted by their formal TOA uncertainties.
In the latter, in quadrature to the white noise uncertainty,  we add a frequency-dependent uncertainty that accounts for the decorrelation of the TOAs. 
We assume that the additional uncertainty can be characterised by a characteristic bandwidth $\Delta \nu_C$ (note that this is not necessarily the scintillation bandwidth) at a frequency $\nu_c$  and has a power-law scaling,
\be
\label{eqn:ism_noise}
\sigma_{\rm uncorr} =   \frac{1}{2 \pi \Delta \nu_C}   \left( \frac{\nu}{\nu_c}\right)^{-Y},
\ee 
where $\Delta \nu_c$ and $Y$ are unknown parameters;  we therefore search for values that minimise the infinite frequency timing error.

\begin{figure*}
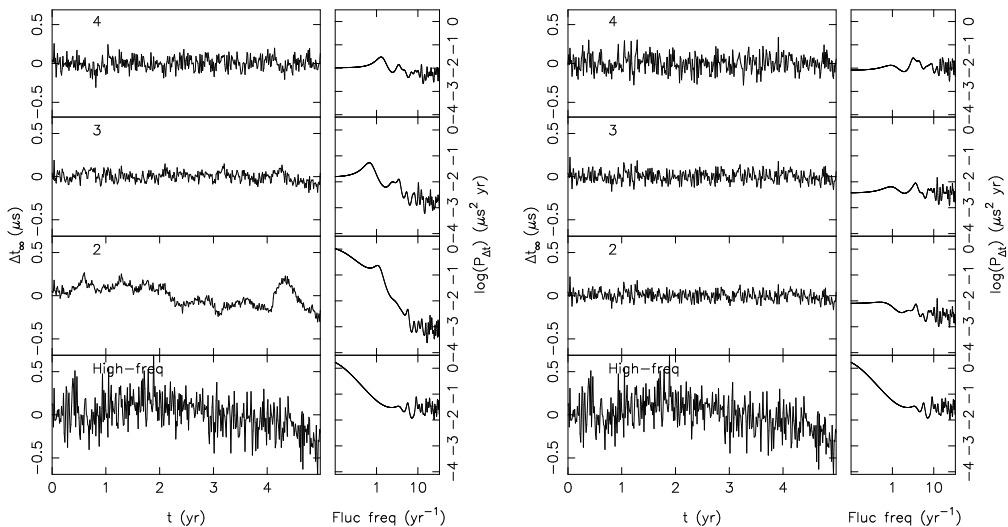

\begin{center} 
\begin{tabular}{cc}
\includegraphics[scale=0.4]{{comp_grand_1937_bad}.eps}  &
\includegraphics[scale=0.4]{{comp_grand_1937_good}.eps} \\
\end{tabular}
\caption{  \label{fig:TOA_corr_pert_kol}  Estimated infinite frequency arrival times $t_\infty$ for a Kolmogorov phase screen with scattering properties like PSR~B1937$+$21.  In the left two panels we show the correction where the TOA uncertainties are used to weight the TOAs.  In the right panels, we show estimates  after accounting for excess scattering noise.   We show the infinite frequency time of arrival $t_\infty$ for three correction schemes (descending from the top row):  four-term  fitting for $\nu^{-2}$, $\nu^{-4}$, and $\nu^{-6.4}$ (4); a three-term approach fitting for $\nu^{-2}$ and $\nu^{-6.4}$ (3); and two-term approach fitting for $\nu^{-2}$ (labeled 2).   We also show the residuals for the highest-frequency arrival times (labeled High-freq) in the lowermost panel. To the right of each time series we show its maximum entropy power spectrum. }  
\end{center}
\end{figure*}

\begin{figure*}
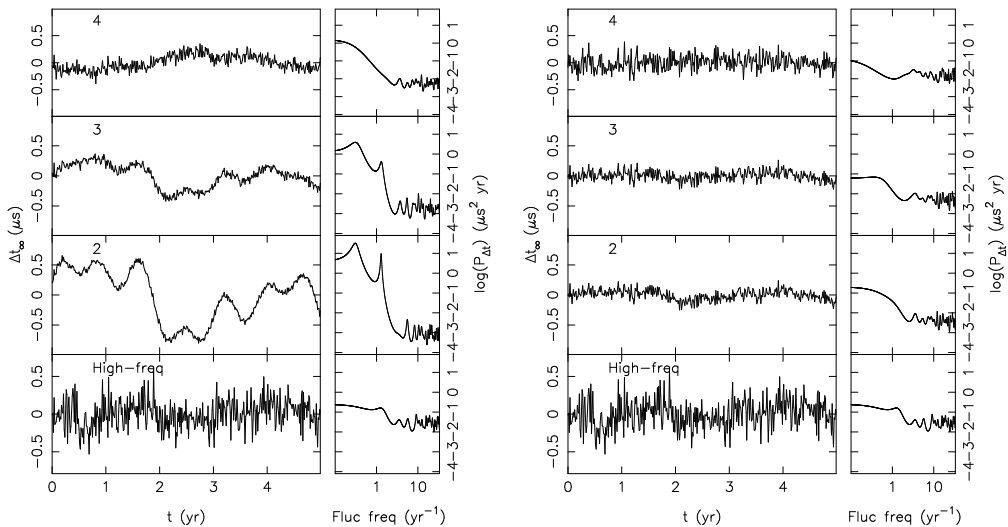

\begin{center} 
\begin{tabular}{cc}
\includegraphics[scale=0.4]{{comp_grand_1937_square_bad}.eps}  &
\includegraphics[scale=0.4]{{comp_grand_1937_square_good}.eps} \\
\end{tabular}
\caption{  \label{fig:TOA_corr_pert_square}   Estimated infinite frequency arrival times $t_\infty$ for a square law phase screen and PSR B1937$-$21 scattering strength.  Panel descriptions are the same as for Figure \ref{fig:TOA_corr_pert_kol}.
}  
\end{center}
\end{figure*}

\begin{figure*}
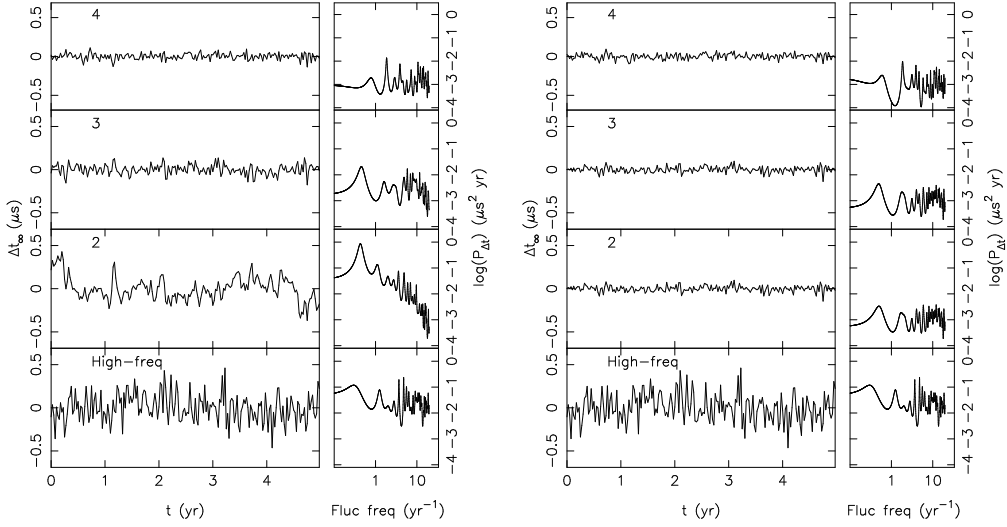

\begin{center} 
\begin{tabular}{cc}
\includegraphics[scale=0.4]{{comp_grand_1713_bad}.eps}  &
\includegraphics[scale=0.4]{{comp_grand_1713_good}.eps} \\
\end{tabular}
\caption{  \label{fig:TOA_corr_pert_1713} Estimated infinite frequency arrival times $t_\infty$ for a Kolmogorov phase screen with scattering properties like PSR~B1713$+$0747.   Panel descriptions are the same as for Figure \ref{fig:TOA_corr_pert_kol}. }  
\end{center}
\end{figure*}

\subsection{Examples and assessment of efficacy}\label{sec:LOS}

In Figures \ref{fig:TOA_corr_pert_kol} to \ref{fig:TOA_corr_pert_1713},  we display the corrected residuals for the same realisations of Kolmogorov and square-law scattering presented in Figures \ref{fig:TOA_pert_k} to \ref{fig:TOA_pert_1713}, representing lines of sight like  PSR B1937$+$21 observed at frequencies between $0.7$~GHz and $3.1$~GHz, and PSR~J1713$+$0747 between $0.2$~GHz and $3.0$~GHz.  

In these cases, we have assumed that the TOA estimation error in a $200$~MHz band close to $1.4$~GHz is 50~ns, and is entirely associated with template-fitting uncertainty.  To determine TOA errors at other frequencies we assumed that the telescope sensitivity is independent of frequency, but that the sky background scales as $S_{\rm sky} \propto \nu^{-2.75}$,  the pulse width scales proportional to $\nu^{-0.3}$ and the pulsar spectrum scales as $\nu^{-2}$.   We do not account for intrinsic jitter noise in estimating the arrival time \cite[][]{2011MNRAS.418.1258O,2012ApJ...761...64S,2014MNRAS.443.1463S} or the finite-scintle effect associated with stochastic pulse broadening due to diffractive scintillation \cite[][]{cs2010}.

In the left series of plots in Figures \ref{fig:TOA_corr_pert_kol} to \ref{fig:TOA_corr_pert_1713}, we show a comparison of the multi-term strategies applied to the TOAs  while retaining  TOA uncertainties that are  their formal white-noise uncertainties. 
To the right of each time series we show its power spectrum.
The estimates of $t_\infty$  derived from the two-term ($\nu^{-2}$) correction scheme show additional noise compared to the three-term and four term strategies.  This noise has a band-pass like power spectrum.  
While three and four-term correction strategies result in  smaller and whiter residual time series compared to two-term mitigation strategies, the additional correction comes at the expense of higher white-noise levels in estimates of $t_\infty$.  

In all cases, the corrected time series are correlated on time scales comparable to the refractive time scale, which is demonstrated through spectral analysis of the time series.  
For reference, we also show the time series of the highest frequency band (3 GHz, labelled high-freq) in Figures  \ref{fig:TOA_corr_pert_kol} to \ref{fig:TOA_corr_pert_1713}.  
The uncorrected residuals show red noise associated with dispersion-measure variations, indicating that for these lines of sight it is insufficient to observe at high-frequency to avoid the effects of the ISM. 

In the right hand series of plots in Figures \ref{fig:TOA_corr_pert_kol} to \ref{fig:TOA_corr_pert_1713}, we show the same mitigation strategies applied, but instead change the weighting by adding (in quadrature) the  ISM-induced uncertainties to the arrival times using Equation (\ref{eqn:ism_noise}).  We find significant improvement in arrival-time estimates using $\nu^{-2}$ correction alone, confirming that down-weighting the low-frequency TOAs (relative to their formal TOA uncertainty) is necessary to reduce errors in estimates of $t_\infty$.

\section{Future studies}\label{sec:future_studies}
 
 The simulations and methods  implemented here can be used to examine a wider range of more complex  scattering geometries, such as anisotropic scattering screens.
 Observations over the last $10$ years indicate that along a large number of lines of sight, density fluctuations are highly anisotropic 
 \cite[][]{2006ApJ...637..346C, 2010ApJ...708..232B}.
 Furthermore some MSPs show evidence for extreme scattering events caused by discrete structures along their lines of sight 
 \cite[][]{2015ApJ...808..113C}
 that cause sudden changes in dispersion, refraction, and diffraction of radio waves. 
  These features can be included by adding additional components to the phase-varying screen or modifying the scattered brightness distribution $B_0(\bm{\theta})$. 

 
Our results would be complemented by  fully diffractive simulations over a wide range of frequencies.     
These diffractive simulations  are necessary to understand the role of ancillary observations of  interstellar propagation such as dynamic spectra and secondary spectra \cite[][]{2005MNRAS.362.1279W,2008ApJ...674L..37H}  in de-biasing arrival-time estimates, and investigating the role of cyclic spectroscopy  \cite[][]{2011MNRAS.416.2821D,2013ApJ...779...99W} in arrival-time correction. 

Our predictions will be tested with higher precision timing observations and observations of MSPs at lower RF.  
For example, the properties of excess noise at low frequencies in a population of MSPs, already evident in PSR~J1643$-$1224 will provide evidence for the propagation effects described here. 
Observations at  very low radio frequencies (such as with the LOFAR telescope) provide the best opportunity to measure these effects where they are the strongest.  
Observations using the International Pulsar Timing Array data sets \cite[][]{2010CQGra..27h4013H,2016MNRAS.458.1267V}, which contain a wide range of frequencies, can be used to identify terms in addition to frequency correlated DM variations.

\section{Conclusions}\label{sec:conclusions}

Without mitigation, the propagation of radio pulses through the  ISM  may impede usage of the highest precision pulsar timing observations for fundamental measurements, including those intending to detect gravitational waves. 
We have assessed the strength of dispersive and refractive interstellar propagation effects on precision pulsar timing observations by simulating TOA perturbations caused by frequency-dependent phase-changing screens.    We then investigated techniques to determine  the infinite frequency time of arrival $t_\infty$ using a variety of scattering screen and observing configurations.   Our main findings are as follows:

1. The ISM induces TOA perturbations that stochastically vary in time, and both deterministically  and stochastically vary in radio frequency.   Through simulation, we  have analysed their strengths and temporal variability.  None of the effects have the steep power spectrum predicted for the GWB, with most showing correlations on week to month time scales.   

2. Multi-frequency observations can be used to correct  propagation effects to a certain extent.    For some nearby, weakly scattered pulsars, sub-$10$~ns rms residuals can be achieved when correcting only for $\nu^{-2}$ perturbations. For other more distant pulsars, correcting for both $\nu^{-2}$ and $\nu^{-6}$ terms produces corrected time series with a higher degree of stationarity.    

3. Optimal observing strategies are highly LOS dependent:   The strengths of the terms depend on the strength and structure of the scattering screen.   As a result, observation and correction  strategies need to be tailored to individual pulsars.  For fixed scattering strength, square law media induce larger and more non-stationary fluctuations in the corrected time series than Kolmogorov media.


4. The best observing strategy depends on the strength of the propagation effects relative to other noise sources:     When the total  observing bandwidth is increased, TOA perturbations become less correlated and correction becomes less effective.  When the total observing bandwidth is narrow, higher order corrections cannot be resolved from the frequency dependent terms.      



5.    Corrected time series contain propagation errors associated with perturbations that vary stochastically with frequency. Even though these effects may be at levels smaller than the white noise, they  affect the sensitivity of a PTA to GWs.  This is further discussed in the context of a measurement model for GW detection (Paper I; \citealt[][]{2012ApJ...750...89C}).   We find that the greatest benefit comes from down-weighting low frequency TOAs 
due to the frequency-dependent stochasticity of the arrival times. 

There is mounting evidence that effects like these are present in MSP observations. 
Recent analysis indicates that the effects of scattering are contributing excess red noise in observations of the millisecond pulsar J1643$-$1224 \cite[][]{iptanoise}. At low frequency ($700-800$~MHz) the pulsar shows stochastic arrival-time variations (in excess of DM variations) with a shallow red-noise spectrum and a strength that is proportional to the pulse broadening time. 
Observations with the PPTA project show that the highest-precision  pulsars show excess noise at the lowest observing frequency \cite{2015Sci...349..1522S}. It is unclear if the noise is instrumental or astrophysical.  If it is astrophysical it could be associated with effects like the ones simulated here. 

\section*{Acknowledgments}

We thank Bill Coles, Rick Jenet, Barney Rickett,  and Dan Stinebring for useful discussions. 
RMS acknowledges travel support from CSIRO through a John Philip early career research award. 
  This work was supported by the U.S. National Science Foundation through grants AST-0807151, PIRE-0968296, and
  NSF Physics Frontier Center award number 1430284.


\appendix

\section{Refracted image shape} \label{sec:app_image_matrices}

In the simulations presented here, the image intensity  $B(\bm{\theta})$ relates  the refractive phase changing screens to refractive and diffractive propagation delay, by weighting the delay through any position by the image intensity expressed in equation (\ref{eqn:delay_intensity}).  
 
 Electron-density fluctuations on large scales refract, focus, and defocus radiation from pulsars, producing variations in the flux density and arrival times of the pulsed radiation.    {\em Large} is defined relative to the refractive scale $D \theta_d$, which is itself larger than the Fresnel scale $\approx \sqrt{\lambda D}$ by the same factor that the Fresnel scale is larger than the diffractive scale $\approx \lambda /\theta_d$ \cite[][]{1990ARAA..28..561R}.  
 While an acceptable model for electron-density variations includes a continuum of length scales, it is useful to separate large and small scales, which we do here.  
 In general, the Kirchhoff diffraction integral for the scalar wave field $E(\bm{x})$ at the observer plane is 
 \be
\label{eqn:KDI}
E(\bm{x})  = - \left[ \frac{i E_p}{ \lambda s(1-s)D} \right]  \int d \bm{x^\prime} \exp\left[i \Phi(\bm{x}, \bm{x^\prime})\right],
\ee
 where $E_p$ is the emitted field, $\Phi$ is the total phase associated with a particular trajectory, and $\bm{x}$ and $\bm{x^\prime}$ are both two-dimensional vectors transverse to the LOS that reside on the observer plane and the screen plane, respectively.  We have denoted the earth-pulsar distance as $D$, the earth-screen distance as $sD$, and the screen-pulsar distance $(1-s)D$.
 The total phase $\Phi = \phi_g  + \phi_r + \phi_d$, is comprised of a geometric term
 \be
 \phi_g (\bm{x}, \bm{x^\prime}) = k \left[ \frac {{x^{\prime}}^2}{sD} + \frac{|\bm{x} - \bm{x^\prime} |^{2}}{(1-s) D} \right],
 \ee
 a refractive term $\phi_r $ from large scale variations, and a diffractive term $\phi_d$ from small scale variations.   
     In the absence of the refractive term, the scattered wave field of a point source is described by its ensemble-average image intensity $B_0(\bm{\theta})$, where $\bm{\theta} = \bm{x^\prime}/sD$.   
 Refraction is included as a paraboloidal perturbation centred on the stationary phase point (SPP) $\bm{\bar{x}^\prime}$:
 \be
 \label{eqn:app_refractive_phase}
&& \phi_r(\bm{x^\prime}) =  \\
&&~~\phi_r(\bm{\bar{x}^\prime}) + \bm{A} \bm{\cdot} (\bm{x^\prime} - \bm{\bar{x}^\prime} ) +  (\bm{x^\prime} - \bm{\bar{x}^\prime} ) \cdot \bm{C} \cdot (\bm{x^\prime} - \bm{\bar{x}^\prime} ), \nonumber  
 \ee
 where $\bm{A}$ is a 2-dimensional vector that describes the phase gradient about the SPP and $\bm{C}$ is a $2\times2$ matrix that describes the curvature about the SPP. 
 
 By consolidating linear and quadratic terms, and solving equation (\ref{eqn:KDI}), the refracted image can be expressed as a function of the undistorted image:
 \be
 B(\bm{\theta}) = B_0[\bm{M}(\bm{\theta}- \bm{\bar{\theta}})],
 \ee
 where $\bm{\bar{\theta}} = \bm{\bar{x}^\prime}/(sD)$ and  $\bm{M} = \bm{U^\dagger} \bm{\gamma_2}^{-1} \bm{U}$, $\bm{\gamma_2}$ is a $2\times2$ diagonal matrix that describes the refractive {\em gains}  along the major and and minor axes of the distortion   and $\bm{U}$ is a $2\times2$ matrix  that diagonalises $\bm{C}$, i.e., a rotation matrix that defines the orientation of the major and minor axes of the paraboloid relative to the coordinate system.    
  
In practice, we do not solve for $\bm{A}$ and $\bm{C}$ at the stationary phase points, but at points on the grid of the phase screen.  
The stationary phase points are located where the total phase gradient is zero:
\be
\label{eqn:app_spp}
\bm{\nabla} \Phi_{\rm tot}(\bm{x^\prime}) = \bm{\nabla} \left( \phi_r + \phi_g \right) =0.
\ee

The location of a SPP can be found by substituting equation (\ref{eqn:app_refractive_phase}) into equation (\ref{eqn:app_spp}).  
Because the resulting equations are linear in $x^\prime$ and $y^\prime$, about each SPP there is a unique solution,
\be
x_r = \frac{cd-ae}{be-c^2}~~{\rm and}~y_r = \frac{ac-bd}{be-c^2},
\ee
where 
\be
a &&= A_x -C_{xx}x_0  -C_{xy} y_0,\\
b &&= \frac{k}{s(1-s) D} + C_{xx},\\
c &&= C_{xy},\\
d &&= A_y - C_{yy} x_0 -C_{xy}y_0,~~{\rm and}\\
e &&= \frac{k}{s(1-s)D} + C_{yy}. 
\ee
In angle, the SPP is located at
\be
\label{eqn:spp_centre}
(\theta_{r,x},  \theta_{r,y})   =(   \frac{x_r}{sD},  \frac{y_r}{sD}).
\ee

The matrix $\bm{M}$ describing the image distortion is
\be
\bm{M} &&= \\ 
&&\left[
\begin{array}{cc} 
G_x \cos^2\alpha + G_y \sin^2 \alpha & ~~~(G_x -G_y) \cos\alpha \sin \alpha\\  
(G_x -G_y) \cos\alpha \sin \alpha & ~~~G_x \sin^2\alpha + G_y \cos^2 \alpha
\end{array}\right], \nonumber
\ee
where  $\alpha$ is the rotation angle of major axis of the curvature relative to the $x$ axis,
\be
\alpha= \frac{1}{2} \tan^{-1} \left[ -2 C_{xy}/(C_{xx}-C_{yy}) \right],
\ee
and  $G_{x,y}$ are the refractive gains along the major and minor axes,
\be
G_x = \left[1 + \frac{s(1-s) D}{k C_{xx}^\prime} \right]^{-1},  G_y = \left[1 + \frac{s(1-s) D}{k C_{yy}^\prime} \right]^{-1}.   
\ee
The curvatures along the major and minor axes are $C_{xx}^\prime$ and $C_{yy}^\prime$:
\be
C_{xx}^\prime = C_{xx} \cos^2 \theta + 2 C_{xy} \cos \theta \sin \theta + C_{yy} \sin^2 \theta \\
C_{yy}^\prime = C_{xx} \sin^2 \theta -  2 C_{xy} \cos \theta \sin \theta + C_{yy} \cos^2 \theta.
\ee

\section{Simulations} \label{sec:simuls_appendix}

In this appendix we describe simulations that were used to model  the perturbations to pulse times of arrival (TOAs) through refractive phase screens.  
The implementation is outlined below and detailed in the following subsections:\\

{\em 1. Define the properties of the scattering screen and the observations} (Appendix \ref{sec:char_scattering}): 
The strength of propagation effects are highly line-of-sight dependent.
Therefore, it is first necessary to define the properties of the scattering screen and both the location and velocity of the pulsar.    
The strengths of the effects are of course also highly dependent on the choice of observational frequencies. \\

{\em 2.   Generate the  reference phase screen $\Phi_{\rm ref}(\bm{x}, \nu_{\rm ref})$.} (Appendix \ref{sec:gen_screen}):    
 As a proxy for electron-density variations that govern radio-wave propagation, the phase screen at the highest observing frequency is  used to construct the phase screens at all observing frequencies because it requires the highest spatial resolution.  \\

{\em 3.  Generate phase screens $\Phi(\bm{x}, \nu)$ for the observing frequencies} (Appendix \ref{sec:norm_screen}):   Relative to the reference phase screen, the other phase screens at lower frequencies have  larger and smoother phase variations. 
 These screens are formed by rescaling the phase and low-pass filtering the reference phase screen $\Phi_{\rm ref}$.   \\

{\em 4.   Calculate propagation delays through the phase screens}  (Section  {\ref{sec:ref_pert} of the main text):  For each time and frequency,  the refractive propagation delays, identified in Section  \ref{sec:analytic_delays},  are calculated.   \\

{\em 5.  Add additional perturbations to the TOAs.}   Other sources of noise are added to the TOAs to properly assess the efficacy of mitigation techniques.   In this paper the only additional noise source considered is  additive white noise.     \\

{\em 6.  Mitigate Propagation Delays} (Section \ref{sec:mitigate} of the main text): We mitigate propagation effects using only  the TOAs themselves.   The properties of the corrected TOAs are quantified by the rms and the spectral properties of the corrected time series.  

\subsection{Characterising the scattering screen} \label{sec:char_scattering}

The phase screens are simulated on a rectangular grid so that at each time $t$ and position $\bm{x}$,  the phase is calculated as $\Phi({\bm r},t;\nu) = \Phi(x_i-s \bm{V_{\rm eff}} t , y_i; \nu)$.  
We define $x$ to be the direction of effective motion  with velocity  $V_{\rm eff}= s V_p$, where $s$ is the fractional distance earth screen distance.
We assume that the   motion of the screen  is negligible, but include the effect of the earth's motion, which can be a significant contribution for the slowest moving MSPs.

In order to estimate the strength of   propagation effects, a number of quantities need to be defined: the geometry of the observer-pulsar-scattering screen system, the nominal strength of the scattering screen, and the observing frequencies. 
 
{\em Calibrating the screen and screen structure:}  
As a proxy for scattering strength, we use the scintillation decorrelation bandwidth $\Delta \nu_d(\nu_{\rm ref})$ at a reference frequency $\nu_{\rm ref}$.  
This quantity is chosen because it is directly observable and often reported in the literature,  and is directly related to other properties of the scattering medium \cite[][]{1986ApJ...310..737C,1986MNRAS.220...19R, 1990ARAA..28..561R}.    

The structure of the fluctuations in the phase screen is characterised  by the power spectrum of the electron density and phase fluctuations $P_{n_e}  \propto P_{\Phi}(\bm{q})$, where $\bm{q}$ is spatial frequency, i.e.,  the Fourier conjugate to screen position $\bm{x}$.     
In most of the simulations we assume that the phase fluctuations follow a power law with power law $P_{\Phi} \propto q^{-\beta}$
over the scales of interest here, which appears to be a reasonable assumption \cite[][]{1990ARAA..28..561R}.
For most lines of sight $\beta$ is between $11/3$ (Kolmogorov) and $4$ (square law).
However, to demonstrate that our simulations are properly behaving, we also simulated a smooth screen by assuming the power spectrum was a one-sided Gaussian ($P_\Phi \propto \exp[ -q^2/2 q_c^2]$).

 
{\em Relevant length scales and numerical screen size:}   
The screen dimensions  are set by the resolution and size requirements of the simulations.  
The relevant length scales are the diffractive length scale $\ell_d(\nu)$, 
the refractive length scale $\ell_r(\nu)$, and the projected distance the pulsar moves along the screen $L_x$.  

At each observing frequency, these length scales are derived from the frequency-dependent decorrelation bandwidth $\Delta \nu_d(\nu)$.   
For power-law media with $\beta$ between $3$ and $4$, the decorrelation bandwidth relative to the reference frequency is
\be
\Delta \nu_d(\nu)   = \Delta \nu_{d,{\rm ref}}  \left(\frac{\nu}{\nu_{\rm ref}} \right)^{2 \beta/(2 - \beta)}.  
\ee

Following the expressions in \cite{1991ApJ...366L..33H}, the diffractive scale on the screen plane is 
\be
\label{eqn:diff_scale_appendix}
\ell_d(\nu) = \pi^{(6-\beta)/(\beta/2-1)}   \sqrt{\frac{s(1-s)c D \Delta \nu_d(\nu)}{2  \nu^2}},
\ee
and the refractive scale on the screen plane is 
\be
\ell_r(\nu) =  \left[ \frac{(\beta-2) }{4(4-\beta)}  \right]^{1/2} \frac{ c s D }{\nu \ell_d(\nu)}.       
\ee

The screen  is set to have grid sizes $\Delta x$ and $\Delta y$ that are both a factor $\mu_i$ smaller than the smallest length scale in the simulation, which is the refractive scale at the highest frequency, $\nu_h$:
\be
\Delta x = \Delta y = \frac{\ell_r(\nu_h)}{\mu_i}. 
\ee

In the direction perpendicular to motion, the size of the  phase screen is set to be larger than the refractive scale at the lowest frequency by a factor $\mu_{o,y}$:
\be
\label{eqn:outer_scale_y}
L_y = \mu_{o, y} \ell_r (\nu_L).
\ee

In the direction of motion, the largest scale is usually the projected distance the pulsar moves across the screen, $s V_p T$, where $T$ is the the total observing span.
\be
L_x  = \mu_{o,x} s V_p T.
\ee
However when the projected distance is small, or the lowest observing frequency is exceptionally low, the refractive scale may be larger.  In this case, $L_x$ follows the form of Equation (\ref{eqn:outer_scale_y}).

To mitigate the effects of periodic boundary conditions associated with Fourier-domain generation of the fluctuations (see discussion below),  
the  length of the screen in the $x$ direction is a factor $\mu_{o,x}$ larger than $L_x$. 

Putting this together, the screen has dimensions
\be
N_x = \frac{L_x}{\Delta x} = \mu_{o,x} \mu_{i}  \frac{s V_p T}{\ell_r(\nu_h)}
\ee
and
\be
N_y = \frac{L_y}{\Delta y} = \mu_{o, y} \mu_{i} \frac{\ell_r(\nu_L)}{\ell_r(\nu_H)}. 
\ee


\subsection{Generating the reference phase screen} \label{sec:gen_screen}

The reference phase screen is generated in the Fourier domain  and then transformed into the spatial domain using a discrete Fourier transform implemented through a fast Fourier transform (FFT) algorithm:  
\be
\label{eqn:phase_e}
\Phi(\bm{r_p}) =  \sum_q \tilde{\Phi}(\bm{q_q}) \exp\left( i  \bm{r_p} \cdot \bm{q_q}\right).
\ee 

The phase screen represents one realisation of  the random medium.  
It is assumed that the screen has a well defined power spectrum\footnote{We note that it is sufficient for the power spectrum to be defined over the observing region. For example, over a finite time (or a finite region of space) many non-stationary processes have well defined ensemble average power spectra, but over infinite time or space they have divergent power spectra.}. 

To generate a single realisation, complex Gaussian white noise $w_q$ is multiplied by an appropriate filter function $F({\bm q})$:
\be
\label{eqn:phase_ft}
\tilde{\Phi}(\bm{q_q}) = w_q F(\bm{q_q}).
\ee
The form of the filter is set by requiring that the power spectrum of the fluctuations follow ensemble average behaviour:
\be
\label{eqn:powspec}
\langle P_\Phi(\bm{q}) \rangle =\langle \tilde{\Phi}({\bm q}) \tilde{\Phi}^*(\bm{q}) \rangle. 
\ee
By substituting  Equation (\ref{eqn:phase_ft}) into Equation (\ref{eqn:powspec}), the  relationship between $P_\Phi$ and $f$ is found to be:
\be
P_{\rm \Phi}(\bm{q}) = \langle w_q^2 \rangle F^2(\bm{q})
\ee

Two types of phase screens were generated.
Smooth screens were generated using a Gaussian shape,
\be
\label{eqn:phase_fluc_smooth}
F(\bm{q}) =   \exp\left( -\frac{q^2 \ell_c ^2}{2}\right),
\ee
where $\ell_c$ is the characteristic length scale for the fluctuations in phase.  For the screen to be smooth, we set $\ell_c$ to be much larger than the refractive scale $\ell_r$.

Turbulent phase screens that show phase fluctuations with power law behaviour $P_\Phi \propto q^{-\beta}$ were also generated.  
In these cases, an appropriate filter is $F(q) = q^{-\beta/2}$.

Because the phase is real-valued in the spatial domain, its Fourier transform $\tilde{\Phi}$ is Hermitian.  
Enforcing  Hermiticity in equation (\ref{eqn:phase_e}), the discrete Fourier transform is 
\be
\label{eqn:phase_cs}
\Phi({\bm x_p}) =  \sum_q F(\bm{q_q}) \left[ a_q \cos( \bm{x_p} \cdot \bm{q_q}) +b_q \sin( \bm{x_p} \cdot \bm{q_q})\right],   
\ee
where $a_q$ and $b_q$ are real random variables with the properties $\langle a_p a_q \rangle = \langle b_p b_q \rangle = \sigma^2\delta_{pq}$, and $\langle a_p b_q \rangle =0 $, where the angled brackets indicate the ensemble-average value of the interior quantity.

\subsection{Generating the multi-frequency phase screens} \label{sec:norm_screen}

For a single realization, all the phase screens at different frequencies are generated from the same template that represents the original electron-density variations.
 In practice, this template phase screen is the highest frequency phase screen because it requires the highest spatial resolution.
 
The strength of the screen at all observing frequencies is  set with the phase structure function:
\be
D_{\rm \Phi}( {\bm b}) = \langle \Phi(\bm{r}) \Phi(\bm{r}+\bm{b}) \rangle.
\ee

For power-law media, the structure function is defined to be
\be
D_\Phi(b) =\pi^2  \left( \frac{b}{\ell_d} \right)^{\beta-2},
\ee
where $\beta=11/3$ for Kolmogorov media and $\beta=4$ for square law media.  
We note that  this definition of the structure function is self consistent with our definition of the diffractive length scale in Equation \ref{eqn:diff_scale_appendix}, but differs from  that in \cite[][]{1990ARAA..28..561R} and our other work 
\cite[][]{2016ApJ...817...16C}.

Using the discrete representation of phase (see Equation \ref{eqn:phase_cs}), the structure function of the simulated phase screen is
\be
D_{\Phi} ( {\bm b}) = 2 \sigma^2 \sum_q F^2({\bm  q_q}) \left[1 - \cos( {\bm q_q} \cdot {\bm b}) \right].
\ee

The screen is normalised on the scale of the smallest resolution element, and  $\sigma$ is chosen such that
\be
\sigma = \left\{ \left( \frac{\pi^2}{2}\right)\frac{D_{\Phi} ( \Delta x/\ell_d)}{ \sum_q F(q)^2   [1 - \cos( {\bm q_q} \cdot {\bm b}) ]}   \right\}^{1/2}.
\ee

To mimic diffractive smoothing of the phase screen, a filter $G(q)$ is applied in the Fourier domain to smooth the phase screen.    The phase screen is smoothed to the refractive scale at a spatial frequency $q_c = 1/ \ell_r(\nu)$ using a Gaussian filter:
\be
G(\bm{q}) = \exp\left[-\frac{q^2}{2 q_c^2} \right].
\ee
Therefore, the phase at frequency $\nu$ is 
\be
\Phi( \bm{r_p}; \nu) = \left(\frac{\nu}{\nu_{\rm ref}}\right)^{-1} \sum_q  G({\bm q_q}) \tilde{\Phi}_r({\bm q_q}) \exp\left[-i {\bm q_q} \cdot \bm{r_p} \right].   
\ee


\end{document}